\documentclass[aps,reprint,twocolumn,pra]{revtex4-1}
\usepackage{amsmath,amssymb}
\usepackage{epsfig}
\usepackage{xcolor,datetime}
\usepackage[pdfstartview=FitV,colorlinks=true,citecolor=blue!80,linkcolor=blue!80]{hyperref}

\newcommand{\dg}{\dagger}
\renewcommand{\a}{\ensuremath{\alpha}}
\newcommand{\p}{\psi}
\newcommand{\la}{\lambda}

\newcommand{\mem}{\text{m}}
\newcommand{\cv}{\text{c,$V$}}
\newcommand{\cc}{\text{c}}
\newcommand{\at}{\text{a}}
\renewcommand{\S}{\ensuremath{\mathcal{S}}}
\newcommand{\s}{\ensuremath{\sigma}}
\newcommand{\zt}{\ensuremath{\zeta}}
\renewcommand{\k}{\ensuremath{\kappa}}
\renewcommand{\b}{\ensuremath{\beta}}
\newcommand{\gam}{\ensuremath{\gamma}}
\newcommand{\oR}{\ensuremath{\omega_R}}
\newcommand{\Om}{\ensuremath{\Omega_\mem}}

\newcommand{\mean}[1]{\ensuremath{\langle #1 \rangle}}
\newcommand{\abs}[1]{\ensuremath{\lvert #1 \rvert}}

\newcommand{\resub}[1]{#1}

\newcommand{\maineqGPE}{\eqref{eq:gpe}}
\newcommand{\maineqHam}{\eqref{eq:ham}}

\begin{document}
\title{Nonequilibrium Quantum Phase Transition in a Hybrid Atom-Optomechanical System}

\author{Niklas Mann$^1$, M. Reza Bakhtiari$^1$, Axel Pelster$^2$, and Michael Thorwart$^1$}
\affiliation{$^1$I.\ Institut f\"ur Theoretische Physik, Universit\"at
Hamburg, Jungiusstra{\ss}e 9, 20355 Hamburg, Germany \\
$^2$Physics Department and Research Center OPTIMAS, Technische Universit\"at
Kaiserslautern, \\
Erwin-Schr\"odinger Stra{\ss}e 46, 67663 Kaiserslautern, Germany }

\date{\currenttime, \today}

\begin{abstract}
We consider a hybrid quantum many-body system formed by a vibrational mode of a nanomembrane, which  interacts optomechanically with light in a cavity, and an ultracold atom gas in the optical lattice of the out-coupled light. The adiabatic elimination of the light field yields an effective Hamiltonian which reveals a competition between the force localizing the atoms and the membrane displacement. At a critical atom-membrane interaction, we find a
nonequilibrium quantum phase transition \resub{from a localized symmetric state of the atom cloud to a shifted symmetry-broken state}, the energy of the lowest collective excitation vanishes, and \resub{a strong atom-membrane entanglement arises.} The effect occurs when the atoms and the membrane are non-resonantly coupled.
\end{abstract}


\maketitle

Hybrid quantum systems combine complementary fields of physics, such as solid-state physics, quantum optics and atom physics, in one set-up. 
Recently, a hybrid atom-optomechanical system \cite{Vogell2013} has been realized experimentally \cite{Jockel2015} in which a single mechanical mode of a nanomembrane in an optical cavity is optically coupled to a far
distant cloud of cold $^{87}$Rb atoms residing in the optical potential of the out-coupled standing wave of the cavity light. When displaced, the membrane experiences  the radiation
pressure force of the cavity light, and in the bad-cavity limit, the field follows the membrane displacement adiabatically. This modulates the light phase which leads to a shaking of the atom gas in the lattice. The nanomechanical motion of the membrane then couples non-resonantly to the collective motion of the atoms.  
The aim  is twofold  \cite{Vogell2013,Jockel2015,Vogell2015,Aspelmeyer2014,Kippenberg2008,Favero2009,Aspelmeyer2010}: The gas can cool the
nanomembrane, and, emergent phenomena of the correlated quantum many-body system are of interest  \cite{Wallquist2009,Hunger2011,Faber2017,Zhong2017,
Meiser2006,Genes2008,Ian2008,Hammerer2009,Hammerer2009-2,Paternostro2010}. 

State-of-the-art optomechanics \cite{Aspelmeyer2014,Kippenberg2008,Favero2009,Aspelmeyer2010} is  nowadays able to realize optical feedback cooling~\cite{Cohadon1999,Kleckner2006} of the mechanical oscillator to its
quantum-mechanical ground state~\cite{Teufel2011,Chan2011}. Yet, the resolved sideband limit allows ground-state cooling only if the oscillator frequency exceeds the photon loss rate in the
cavity~\cite{Neuhauser1978,Marquardt2007,Schliesser2008}. Hence, cooling a macroscopic \textit{low-frequency} nanomembrane close to its ground state is so far not possible.
One promising alternative \cite{Vogell2013,Vogell2015} is to utilize an ultracold atom  
gas, which has been demonstrated recently \cite{Jockel2015} by 
sympathetic cooling down to 650 mK. Current investigations aim to a coherent state transfer of robust quantum entanglement  \cite{Hammerer2009}. 

Apart from cooling the nanomembrane, an interesting fundamental feature is the collective quantum-many body behavior of the hybrid system. For instance, the atom-atom interaction can in principle be
coherently modulated by the back-action of the cavity light on the nanooscillator. 
By this, a long-range interaction emerges which resembles that of a dipolar Bose-Einstein condensate (BEC)~\cite{Lahaye2009}. In fact, a simpler hybrid quantum  many-body system has also been implemented in the
form of a
BEC in an optical lattice inside a transversely pumped optical cavity. A Dicke quantum phase transition between a normal phase and a self-organized superradiant
phase occurs~\cite{Nagy2008,Maschler2008,Baumann2010,Bakhtiari2015,Klinder2015}. Moreover,  optical bistability \cite{Gupta2007,Ritter2009}, a roton-type softening in the atomic  dispersion
relation~\cite{Nagy2008,Oztop2012,Mottl2012,Leonard2017} and optomechanical Bloch oscillations~\cite{Kessler2016} were uncovered. 
\resub{Similar effects occur also for polarizable and thermal particles in a cavity at finite temperature~\cite{Asboth2005,Lu2007,Salzburger2009}.}
\begin{figure}[b]
  \begin{center}
\includegraphics*[clip=true, trim=0 0 0 0, width=0.9\columnwidth]{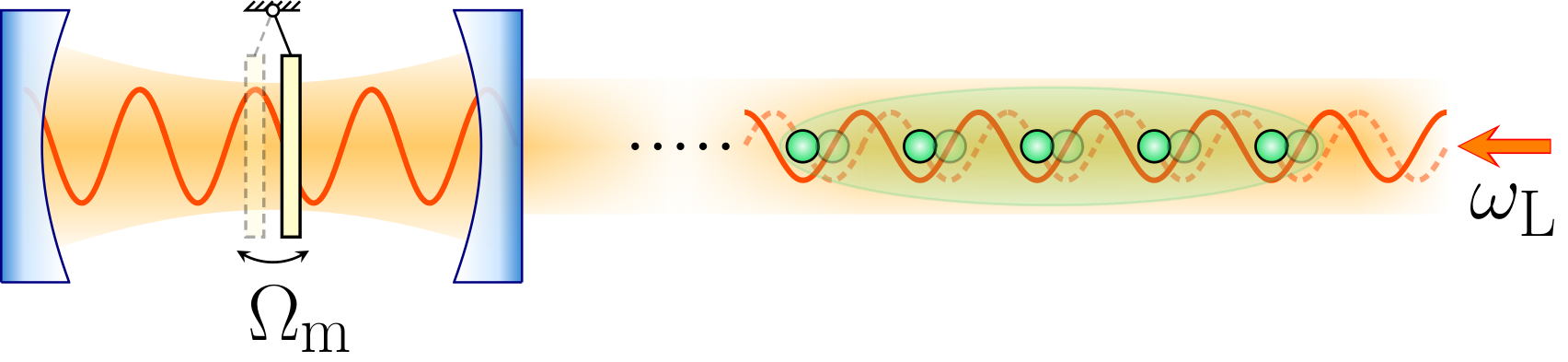} 
\caption{Sketch of the hybrid system. A nanomechanical membrane in an optical cavity is optically 
coupled to the vibrational motion of a distant atom gas.
}\label{figSk}
\end{center}
\end{figure}

In this work, motivated by recent experiments \cite{Jockel2015, Faber2017}, we study a hybrid atom-optomechanical setup in the form of a ``membrane-in-the-middle''
cavity \cite{Wallquist2009,Hunger2011,Vogell2013,Jockel2015,Faber2017,Zhong2017}, see Fig.\ \ref{figSk}.  
Importantly, the light-mediated coupling between the atoms and the membrane is non-resonant here. 
We 
include the full lattice potential  and also the atomic interaction in the gas on the mean-field level. The 
numerical solution of the generalized Gross-Pitaevskii equation confirms the validity of an analytic approach based on a Gaussian condensate profile. 
Tuning the atom-membrane coupling by changing the laser intensity, a nonequilibrium quantum phase transition (NQPT) occurs 
between a localized symmetric state and a symmetry-broken quantum many-body state with a shifted cloud-membrane configuration. It is fueled by the competition of the lattice, trying to localize the atoms 
at the minima, and the  membrane displacement which tries to shake the atoms. 
Near the quantum critical point, the energy of the lowest collective excitation mode vanishes and the order parameter of the symmetry-broken state becomes non-zero\resub{, leading to a substantial atom-membrane entanglement}.
The mode-softening is accompanied by a roton-type bifurcation of the decay rate of the collective eigenmodes. 
Indeed, an instability and collective self-oscillations of a coupled atom-membrane device have been reported recently ~\cite{Faber2017}.

\section{Model}

We consider a single mechanical mode of a nanomembrane with frequency $\Om$ placed in a low-finesse optical cavity. The outcoupled light forms an optical lattice in which a BEC is placed. 
In a quasistatic picture, a finite displacement of the membrane changes the position of the lattice sites, leading to a linear displacement force on the atoms which induces transitions to higher motional bands.
A back-action of the atomic motion on the membrane is induced by a displacement of their center-of-mass position,  which, again, redistributes the photons in the propagating beams. Consequently, the light field
 inside the cavity changes, which alters the radiation pressure on the membrane. 
To achieve a sizable atom-membrane coupling, the typical energy scale of the BEC has to match the membrane frequency. This can be controlled by the light intensity which determines the lattice depth. The set-up
is sketched in Fig.~\ref{figSk} and modeled by a standard Hamiltonian, which describes the atom-membrane coupling directly
\cite{Vogell2013} (see \ref{App1}). \resub{In the bad-cavity limit, strong photon dissipation allows us to adiabatically eliminate the light field in a Born-Markov approximation \cite{Vogell2013} which} yields the effective Hamiltonian
\begin{align}\label{eq:ham}
 H &= \int \! dz \Psi^\dg(z)\Bigl[-\oR\partial_z^2+V\sin^2(z) + \frac{g}{2}\Psi^\dg(z)\Psi(z)\Bigr]\Psi(z) \nonumber\\
 & + \Om a^\dagger a - \la(a^\dg+a)\int dz \Psi^\dg(z)\sin(2z)\Psi(z)\,.
\end{align}
Here, $\oR=\omega_\text{L}^2/2m$ is the recoil frequency of an atom with mass $m$, $V$ the optical lattice depth, and $\omega_\text{L}$ the laser frequency. The last term describes the effective atom-membrane coupling
with strength $\lambda$. 
$a(a^\dg)$ and $\Psi(\Psi^\dg)$ are the bosonic annihilation (creation) operators of the membrane and bosons. 
Moreover, we have introduced a local atom-atom interaction with strength $g$ 
and neglected long-range interaction, 
which is generated by the photon field. This is justified when the laser frequency is far detuned from the closest atomic transition.

In the condensate regime, a large fraction of the atoms occupies the ground state. Here, we consider weakly interacting atoms that are also weakly coupled to the membrane.
Thus, when $g,\la\ll\oR,\Om$, the field operator $\Psi(z)$ can be approximated by a complex function $\psi(z)$ according to $\Psi(z)\simeq\sqrt{N}\psi(z)$, where $N$ denotes the number of atoms. To describe the 
 dynamics, we use the mean-field Lagrangian density associated to the Hamiltonian and given in Eq.\ \eqref{eq:La} with the complex number $\mean{a}/\sqrt{N}=\a=\a'+i\a''$ and the volume $\mathcal{V}$. 
We restrict the problem to a single lattice site, i.e., $\int dz\rightarrow\int_{-\pi/2}^{\pi/2}dz$ and $\mathcal{V}=\pi$, and use periodic boundary conditions. Then, we describe the dynamics analytically with a Gaussian ansatz for the condensate wave function
and, in parallel, solve the generalized Gross-Pitaevskii equation (GPE) without further approximation.  
The Euler-Lagrange equations yield
\begin{gather}
 i\partial_t\psi \!= \![V\sin^2(z) -\oR\partial_z^2+ gN\abs{\psi}^2-2\sqrt{N}\la \a'\sin(2z)]\psi,\nonumber\\
i\partial_t \a =({\Om-i\gamma})\a-\sqrt{N}\la\int dz\sin(2z)\abs{\psi}^2\,,
\label{eq:gpe}\end{gather}
where we have introduced a phenomenological damping of the mechanical mode with a rate $\gam$. This is due to finite losses caused by the clamping of the membrane as well as the radiation pressure. 

From Eq.\ (\ref{eq:gpe}), we see that the two potential 
contributions $V\sin^2(z)$ and $\sqrt{N}\la(\a+\a^*)\sin(2z)$ can dynamically compete with each other, depending  on the back-action of the membrane on the atoms and, thus, on the collective 
behavior of the atoms. This competition yields to the formation of two different stable phases and 
a NQPT. It is manifest in a change of the center-of-mass position of the condensate, or the membrane displacement, equivalently. \resub{Formal similarities to the NQPT in the Dicke-Hubbard model \cite{Bakhtiari2015,Klinder2015} exist. There, however, 
 a self-organized symmetry-broken checkerboard lattice occupation is formed above a 
critical transverse pump strength which induces a coherent light scattering into the longitudinal cavity mode   \cite{Maschler2008,Nagy2008}.
In the present set-up, the spatial periodicity of the optical lattice remains unchanged, and symmetry breaking is manifest in a global collective shift of the potential.}

To describe a realistic physical set-up, 
we consider a membrane with $\Om=100\,\oR $, which corresponds to a frequency of several hundred kHz. 
\resub{Here, we describe the condensate profile by a Gaussian \cite{pelster}}
\begin{equation}
\psi(z,t)=\left[\frac{1}{\pi\s(t)^2}\right]^{1/4} e^{{-\tfrac{[z-\zt(t)]^2}{2\s(t)^2}+i\k(t) z + i\b(t) z^2}}\,,
\end{equation}
with a time-dependent width $\s (t)$, centered at the position $\zt (t)$, and the corresponding phases $\b(t)$ and $\k(t)$.
\resub{For an accurate description, we consider $V\gtrsim10\,\oR$ and $Ng\ll V$.}
To find the equations of motion of these variational parameters, we determine the lowest cumulants of the condensate probability distribution whose dynamics is
described by the generalized GPE \eqref{eq:gpe}. Thus, we
%
 multiply Eq.\ \eqref{eq:gpe} (i) by $\psi^*(z,t)(z-\zt)$ and integrate over $z$, and, likewise, (ii)
by $\psi^*(z,t)\left[(z-\zt)^2-\tfrac{\s^2}{2}\right]$ and integrate over $z$.
%
This yields four linearly independent equations for the variational parameters, two of which are  $\dot\zt=2\oR(\k+2\b\zt)$ and $\dot\s=4\oR\b\s$.
With these, we find
%
%
%
%
\begin{figure*}
  \begin{center}
\includegraphics*[clip=true, trim=20 3 0 2, width=0.92\textwidth]{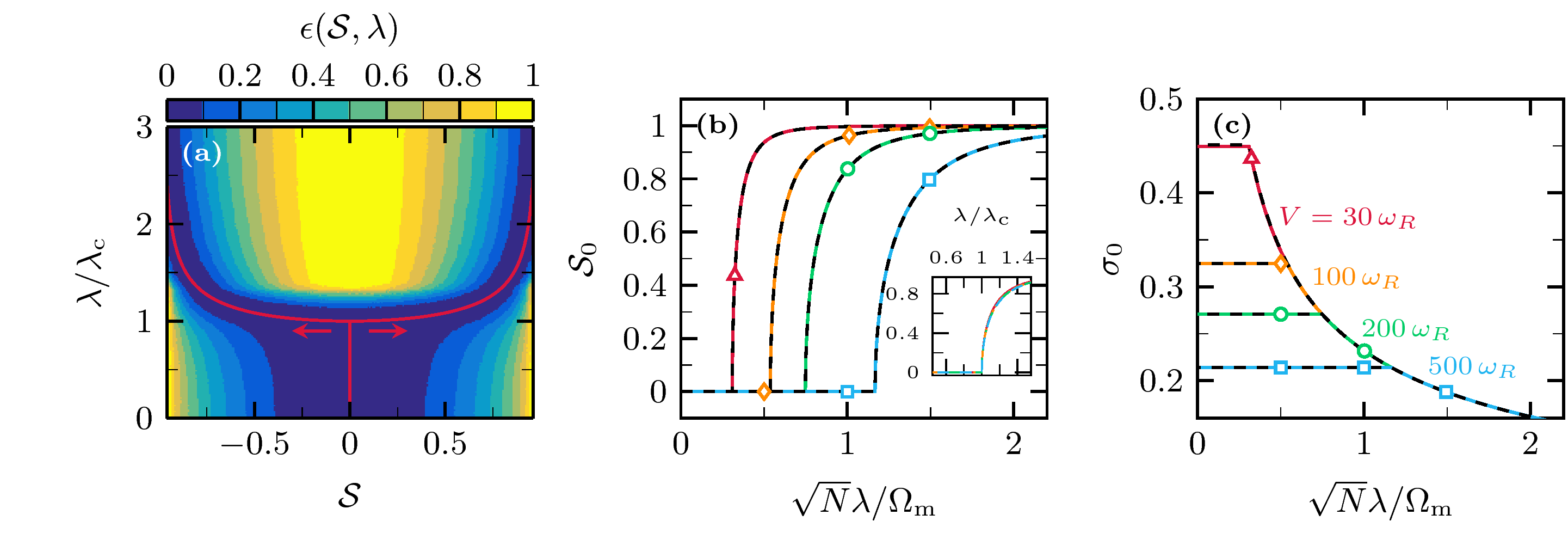}
\caption{(a) Normalized potential energy surface $\epsilon(\S,\la)$ 
  as a function of $\S$ and $\la$ for $V=200\,\oR$. The red line indicates the minimum $\epsilon(\S_0,\la)=0$. (b) Positive value of $\S_0$ and (c) condensate width $\s_0$ as a function of $\la$ for different values $V$ as
  indicated in (c). The dashed curves show the GPE results for comparison. The inset in (b) shows the order parameter $\S_0$ as a function of $\la/\la_\cc$ for which all data points collapse to a single curve. For all panels, we have used $g=0$, $\Om=100\,\oR$ and $\gamma=20\,\oR$.}\label{fig1}
\end{center}
\end{figure*}
\begin{align}
{2}{\Om^{-1}}[\ddot\a'+2\gam\dot\a']&=-{\partial_{\a'} E}\,,\nonumber\\
({2\oR})^{-1}\ddot\zt&=-\partial_\zt E\,,\label{eq:LEoM}\\
({4\oR})^{-1}\ddot{\s}&= -{\partial_\s E}\, , \nonumber
\end{align}
where the potential energy $E=-\int dz \mathcal{L}\rvert_{\dot\a=\dot\psi=0}$ reads
\begin{equation}
\label{energy}
E=\frac{gN}{\sqrt{8\pi}\s}-\frac{V\sqrt{1-\S^2}+4\sqrt{N}\la\a'\S}{2\exp(\s^2)}+\frac{\oR}{2\s^2}+\tilde{\Omega}_\mem\a'^2
\end{equation}
with the effective frequency $\tilde{\Omega}_\mem=\Om+\gam^2/\Om$.
Importantly, we have defined the order parameter $\S=\sin(2\zt)$ 
 of the NQPT, which describes the center-of-mass position of the condensate.
\section{Quantum Phase Transition in the Mean-Field Regime}
Due to the mechanical damping, the combined system will eventually equilibrate. 
The steady state is characterized by those values $\a_0',\s_0,\S_0$ which minimize the potential energy functional 
$E(\a',\s,\S)$. Indeed, by setting all time-derivatives in Eq.\ \eqref{eq:LEoM} to zero and using  Eq.\ (\ref{energy}), we find the 
relation $\sqrt{N}\la \S_0= \tilde{\Omega}_\mem \a_0'e^{\s_0^2}$
, so that the equilibrium width $\s_0$ and order
parameter $\S_0$ solve the coupled equations
$(1-\S_0^2)^{1/2}[\oR+gN\s_0/\sqrt{8\pi}]=V\s_0^4 e^{-\s_0^2}$ and
$\S_0[N\la^2(1-\S_0^2)^{1/2}-N\la_\cv^2e^{\s_0^2}]=0$
with $N\la_\cv^2= \tilde{\Omega}_\mem V/4$.

For a qualitative understanding of the role of increasing $\la$, we define the potential energy surface as a 
function of a single variable, i.e., either $\s$ or $\S$ (or $\a'$).
For instance, $E(\s)\equiv E(\a_0'(\s),\s,\S_0(\s))$ exhibits only a single minimum 
for $\s>0$. Interestingly enough, as a function of the control parameter~$\la$, 
the energy  $E(\S)$ 
has either one stable state or two minima.
This is visualized by the normalized potential energy surface $\epsilon(\S,\la)=
\tfrac{E(\S)-E(\S_0)}{\text{max}\{E(\S)-E(\S_0)\}}$ in Fig.\ \ref{fig1}(a).
The red curve marks the configuration of minimal energy 
$\epsilon(\S_0,\la)=0$. There exists a critical coupling $\la_\cc$, such that
for smaller values $\la<\la_\cc$, the energy surface 
forms a single potential well, whereas for $\la>\la_\cc$,  it becomes a double 
well potential with a local maximum at $\S=0$.

The order parameter as a function of the atom-membrane coupling $\la$ is shown in 
Fig.~\ref{fig1}(b) for different values of the lattice depth. The solid curves show the results of the analytical approach, whereas the dashed lines refer to the numerical solution of the full GPE.
For small values $\la$, the condensate is symmetrically located 
around the lattice minima $\zt_0=j\pi$ with $j\in\mathbb{Z}$, so the order parameter vanishes, $\S_0=0$.
Consequently, the membrane displacement $\a_0'\sim\S_0$ vanishes. The NQPT then occurs at a critical coupling $\la_\cc$, which follows from solving the implicit equation
\begin{equation}
\label{critical}
\oR + \frac{gN}{\sqrt{8\pi}}\,\sqrt{2 \, \ln \frac{\la_\cc}{\la_\cv}} = 4 V \left( \frac{\la_\cv}{\la_\cc} \ln \frac{\la_\cc}{\la_\cv} \right)^2\,.
\end{equation}
Above $\la_\cc$, the atoms start to move away from the positions $j\pi$ to the 
displaced lattice minima. 
The order parameter becomes finite: $\S_0=\pm\Theta(\la-\la_\cc)\sqrt{1-({\la_\cv}/{\la})^4 e^{2\s_0^2}}$
\resub{ and can be scaled to a single curve as shown in the inset of Fig.~\ref{fig1}(b).}
The condensate width $\s_0(\la)$ is shown in Fig.~\ref{fig1}(c) and is 
independent of $\la$ below $\la_\cc$, whereas it decreases in good approximation with $\sim1/\sqrt{\la}$ above $\la_\cc$.
In accordance with an expansion of the energy surface with respect to the order parameter, all these observables show within our mean-field treatment that the hybrid system undergoes a second order NQPT. 
\resub{In contrast to the superradiant phase of the Dicke phase transition, a global displacement of the lattice minima and membrane is observed and the lattice periodicity is not changed.}
\section{Collective Excitation Modes}
Solving the complete set of equations of motion~\eqref{eq:LEoM} is challenging, but their linearized forms give  already an insight into the collective excitation energies. We consider small deviations
from the stationary state $(\a_0',\s_0,\S_0(\text{or }\zt_0))$ in the form of $\a'(t)=\a_0'+\delta\a'(t)$, $\s(t)=\s_0+\delta\s(t)$ and $\zt(t)=\zt_0+\delta\zt(t)$ and linearize the equations of motion in the deviations. We find the Eqs.\ \eqref{eq:LEoM-st}.
%
%
Interestingly, the oscillation frequencies also indicate the NQPT. 
Below $\la_c$, the bare frequency $\omega_\zt=\sqrt{4\oR{}V}e^{-\s_0^2/2}$ of the $\zt$-mode is 
independent of $\la$,  whereas, above $\la_c$, it grows linearly in $\la$ according to 
$\omega_\zt=\sqrt{4\oR{}V}e^{-\s_0^2}\la/\la_\cc$.
\begin{figure}
\begin{center}
\includegraphics*[clip=true, trim=20 4 0 0, width=0.83\columnwidth]{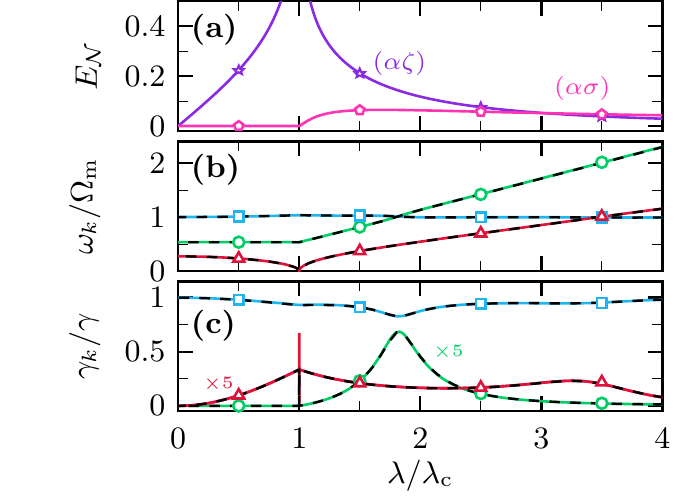} 
\caption{(a) Atom-membrane entanglement by the logarithmic negativity of the ground state ($\gamma=0$) between membrane and either atom displacement (purple) or condensate width (pink), (b) frequencies of collective excitations, and, (c) decay rates of the collective eigenmodes (some curves are scaled by the factors indicated).
  Curves in blue, green, and red correspond to the membrane mode, the condensate width mode, the atomic displacement excitation, correspondingly. The dashed curves show the GPE
results. The parameters are $V=200\,\oR$, $\Om=100\,\oR$, $\gam=20\,\oR$ and $g=0$.}\label{fig2}
\end{center}
\end{figure}

In addition, the eigenmodes can be determined (for details see \ref{App3}) from the differential equations in the vector-matrix-form $\dot{\boldsymbol{x}}=\boldsymbol{M}\boldsymbol{x}$. 
The eigenvalues $\nu_k=i\omega_k-\gam_k$ of $\boldsymbol{M}$ define the eigenfrequencies $\omega_k$ and the decay rates $\gam_k$.
Likewise, we estimate the eigenmodes via the GPE by considering small deviations from the ground state 
according to $\psi(z,t)=e^{i\mu t}[\psi_0(z) + \delta\psi(z,t)]$ and $\a(t)=\a_0+\delta\a(t)$.
Linearization with respect to the deviations results in a differential equation of the form $\dot{\boldsymbol{w}}(z)=\boldsymbol{M}_{\rm GP}\boldsymbol{w}(z)$, where the eigenvalues of $\boldsymbol{M}_{\rm GP}$ provide the
eigenfrequencies and the decay rates (see also \ref{App4}).

The eigenfrequencies of the collective excitations without interatomic collisions are shown in 
Fig.\ \ref{fig2}(b) and the corresponding decay rates in Fig.\ \ref{fig2}(c) as a function of the atom-membrane coupling strength \resub{(Figs.\ \ref{fig3old} (b) and (d) show zooms to the critical region)}.
The dashed lines show the frequency (rate) calculated in the GPE approach, whereas the solid lines refer to 
the analytical results.
Approaching the critical coupling, the lowest excitation frequency (red, triangles) in Fig.~\ref{fig2}(b) decreases with a roton-type behavior according to $\sim\sqrt{1-\la^2/\la_\cc^2}$.
At the same time, the corresponding decay rate increases up to a maximum at~$\la_\cc$.
In a narrow range $\Delta\la\simeq0.1\,\oR$ around $\la_\cc$, a bifurcation of the decay rate can be 
 observed, whereas the lowest excitation frequency is constantly zero.
Such a behavior is well known from atomic ensembles with long-range interactions \cite{Mottl2012,Leonard2017,Santos2003,Giovanazzi2004} which, in the present case, are mediated by the membrane.
Indeed, adiabatically eliminating the membrane mode introduces a long-range interaction potential that takes the form $G(z,z')=G_0\sin(2z)\sin(2z')$ with $G_0=-2\la^2 /  \tilde{\Omega}_\mem $.
\resub{Moreover, the ground state ($\gamma=0$) of the collective modes is a three-mode squeezed state, see Eq.\ \eqref{eq:ham2}. This generates a strong atom-membrane entanglement close to the critical point.
This behavior is manifest in a rising logarithmic negativity $E_\mathcal{N}$~\cite{Adesso2004,Nagy2011,Arani2016} at the critical point which is shown in Fig.~\ref{fig2}(a). For $\gamma=0$, it diverges there while it 
is expected to be finite for $\gamma>0$ \cite{Adesso2004}. 
}
Finally, we note that although all figures refer to the case $g=0$, 
no qualitative differences occur for weakly interacting atoms (see \ref{App6}).

\section{Experimental realization} 
\begin{figure}
\begin{center}
\includegraphics*[clip=true, trim=20 2 0 0, width=0.92\columnwidth]{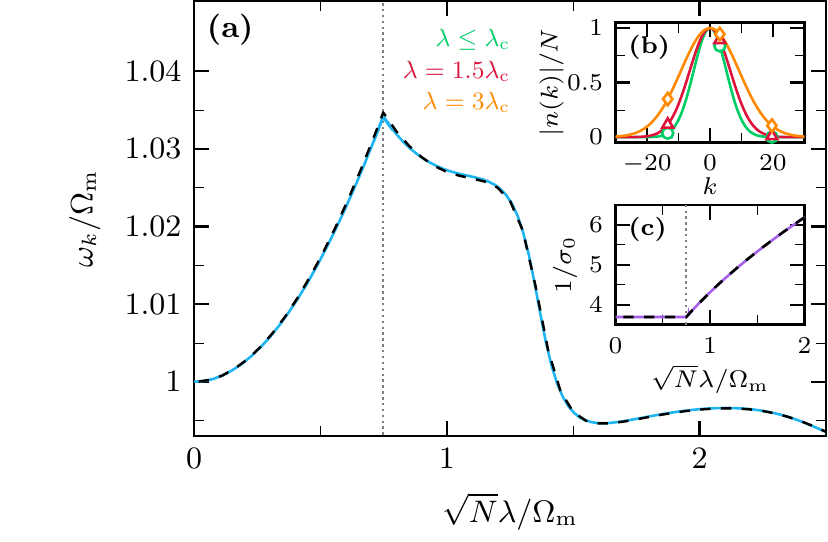}
\caption{(a) Membrane excitation frequency, and, (b) momentum distribution $\abs{n(k)}$ of the atoms in a single well. 
(c) Width of $\abs{n(k)}$ in dependence of $\la$. The dotted vertical line indicates $\la_\cc$. The parameters are $V=200\,\oR$, $\Om=100\,\oR$, $\gam=20\,\oR$ and $g=0$.}\label{fig4}
\end{center}
\end{figure}

An experimental observation is possible in existing set-ups, e.g., in Ref.\ \cite{Zhong2017}. \resub{Current optical lattices with $V\simeq2000\,\oR$ readily achieve a resonant coupling \ref{App5}, i.e., $\omega_\zt\simeq\Om$, with $\sqrt{N}\la\simeq3\,\oR$. The fact that the effective atom-membrane coupling in our scheme does not have to be resonant facilitates the realization. 
 For instance, by loading the atoms in a lattice with $V=30\,\oR$, $\sqrt{N}\la_\cc\simeq30\,\oR$ can be reached by tuning the laser power and cavity finesse~\cite{Vogell2013}.
}
Moreover, an independent tuning of $\la$ can be achieved by applying a second laser which is slightly misaligned with the first one and  
 which generates an optical lattice of the same periodicity but shifted by $\pi/2$.
\resub{The membrane eigenfrequency shown in Fig.\ \ref{fig4}(a) can readily be measured  spectroscopically with a  relative precision of much below $1\%$, so that the cusp at $\la_\cc$ will be clearly resolvable. 
In addition, the NQPT can also be detected via the momentum distribution 
of the atoms shown in Fig.\ \ref{fig4}(b), together with its width for varying $\la$ in Fig.\ \ref{fig4}(c). Below $\la_\cc$, the width is constant, while it increases monotonically for $\la > \la_\cc$.}

\section{Conclusions}
We have shown that a hybrid atom-optomechanical system possesses a nonequilibrium quantum phase transition between phases of different collective behavior. Based on a Gross-Pitaevskii like mean-field approach,
the steady state  of an ultracold atomic condensate in an optical lattice,  whose motion is non-resonantly coupled to a single mechanical vibrational mode of a spatially distant membrane, has been analyzed. The coupling between 
both parts occurs via the light field of a common laser.  
Below the critical effective atom-membrane coupling $\la_\cc$, the atoms in the combined atom-membrane ground state are symmetrically distributed around their lattice minima. 
At the quantum critical point, a nonequilibrium quantum phase transition to a symmetry-broken state occurs in which the atomic center-of-mass and membrane displacements are all either positive or negative.
Near the NQPT, the lowest excitation mode shows roton-type characteristics in the excitation frequency, a mode softening and a bifurcation of the decay rate, accompanied by a strong atom-membrane entanglement. \resub{A potential application could be to measure the atom momentum fluctuations non-destructively by measuring the fluctuations of the membrane displacement.}  

\begin{acknowledgments}
This work was supported by the Deutsche Forschungsgemeinschaft (N.M.), by the 
DFG Collaborative Research Center 928 ''Light-induced dynamics and control of correlated quantum systems'' (M.R.B.) and the DFG Collaborative Research Center/TransRegio 185 "Open System Control of Atomic and
Photonic Matter" (A.P.). We thank Christoph Becker, Alexander Schwarz and Hai Zhong for useful discussions on experimental set-ups.
\end{acknowledgments}

\appendix

\section{Effective Hamiltonian of the hybrid system}
We start from the Hamiltonian~\cite{Vogell2013}\label{App1}
\begin{equation}
 H = H_\mem + H_\at + H_\text{l} + H_\text{m-l} + H_\text{a-l} \,.
\end{equation}
The first three terms describe the free time evolution of the single mechanical mode of the nanomembrane, the atomic condensate and the light field, respectively.
In addition, the coupling between the light field  and the atoms and between the light field and the membrane is denoted by  $H_\text{a-l}$ and $H_\text{m-l}$, respectively. The single membrane mode is modeled as a harmonic oscillator with frequency $\Om$ with the Hamiltonian 
\begin{equation}
H_\mem = \Om a^\dag{}a\,,
\end{equation}
and the atomic condensate by the usual many-body Hamiltonian in second quantization
\begin{equation}
H_\at=-\oR\int\! dz\,\Psi^\dag(z)\partial_z^2\Psi(z) +\frac{g}{2}\int\! dz\, \abs{\Psi(z)}^4\,,
\end{equation}
where $\oR=\omega_\text{L}^2/2m$ is the recoil frequency, $\omega_\text{L}$ the laser frequency and $g$ the local interaction strength. The light modes are included over a bandwidth $2\theta$ around the laser frequency by the Hamiltonian 
\begin{equation}
H_\text{l} = \int_{\omega_\text{L}-\theta}^{\omega_\text{L}+\theta} \!d\omega\, \Delta_\omega b^\dag_\omega b_\omega\,,
\end{equation}
with $\Delta_\omega=\omega-\omega_L$.
The above defined operators fulfill the usual bosonic commutation relations, i.e., $[a,a^\dag]=1$, $[b_\omega,b^\dag_\nu]=\delta_{\omega\nu}$ and $[\Psi(z),\Psi^\dag(z')]=\delta(z-z')$. 

By a linear replacement $b_\omega\rightarrow b_\omega + \delta(\omega-\omega_\text{L})e^{-i\omega_\text{L}t}\alpha_\text{L}$ with field strength $\alpha_\text{L}$, the external laser drive is included.
Then, linearizing around $\alpha_\text{L}$ leads to the membrane-light coupling 
\begin{equation}
 H_\text{m-l}= \la_\mem (a+a^\dag) \int\!\frac{d\omega}{\sqrt{2\pi}}\,(b_\omega+b^\dag_\omega)\,,
\end{equation}
with the coupling strength $\la_\mem$.
On the other hand, the laser field induces a Stark shift in the ground state of the atoms.
Consequently, the atom-light coupling is described by the interaction Hamiltonian
\begin{align}
&H_\text{a-l} = V \int \!dz \,\Psi^\dag(z)\sin^2(z)\Psi(z)\\
&+\la_\at \int \!\frac{d\omega}{\sqrt{2\pi}}\,(b_\omega+b^\dag_\omega) \int\!dz\, \Psi^\dag(z)\sin(z)\sin(\tfrac{\omega}{\omega_\text{L}}z)\Psi(z)\, , \nonumber
\end{align}
with the optical lattice depth $V$ and the atom-light coupling strength $\la_\at$.
To obtain an effective Hamiltonian for the atom-membrane coupling, the idea is the following. First, we derive the Heisenberg equation for the field quadrature $x_\omega=(b_\omega+b_\omega^\dag)/\sqrt{2}$
\begin{align}\begin{split}
\ddot{x}_\omega &+ \Delta_\omega^2 x_\omega = -\frac{\la_\mem\Delta_\omega}{\sqrt{\pi}}(a+a^\dag)\\&-\frac{\la_\at\Delta_\omega}{\sqrt{\pi}}\int dz \Psi^\dag(z)\sin(z)\sin(\tfrac{\omega}{\omega_\text{L}}z)\Psi(z)\,.
\end{split}\label{eq:xw}\end{align}
Then, we insert the formal solution of \eqref{eq:xw} in the equations of motion for $a$ and $\Psi(z)$.
The emerging integrals over the field modes $\omega$ are of the form
\begin{equation}
\int \! d\omega \sin(\tfrac{\Delta_\omega}{\omega_L} \tau) \sin(\tfrac{\omega}{\omega_L}z)\underset{\theta\rightarrow\infty}{\longrightarrow} \pi\cos(z)[\delta(\tau-z)-\delta(\tau+z)]\,.
\end{equation}
Finally, neglecting the advanced term and using the relation $\cos(z)\sin(z)=\sin(2z)/2$, one may insert the result in the equations of motion for $a$, $\Psi$ and read off the effective Hamiltonian given in Eq.~\maineqHam~of the main text with the effective coupling $\la=\la_\at\la_\mem/2$. It is important to realize that 
the derivation exploits the assumption of a strong photon loss in the cavity, see Ref.\ \cite{Vogell2013}. This corresponds to a rapid approach of the light field to its steady state and justifies a Markov approximation. 
%
%

%
%
\section{Lagrangian density}\label{App2}
The Lagrangian density of the hybrid system can be directly inferred from the effective Hamiltonian in Eq.\ \eqref{eq:ham} in the main paper. It reads
\begin{widetext}\begin{equation}
\mathcal{L} = \frac{1}{\mathcal{V}}\Bigl[\frac{i}{2}(\dot{\a}\a^*-\a\dot{\a}^*)-{\Om} \a^*\a\Bigr] + \frac{i}{2}(\dot{\psi}\psi^*-\psi\dot{\psi})-\Bigl[\oR\abs{\partial_z\psi}^2+V\sin^2(z)\abs{\psi}^2
+\frac{gN}{2}\abs{\psi}^4-\sqrt{\la}(\a+\a^*)\sin(2z)\abs{\psi}^2\Bigr]\,,\label{eq:La}
\end{equation}\end{widetext}
with the complex number $\mean{a}/\sqrt{N}=\a=\a'+i\a''$ and the volume $\mathcal{V}$. 
{
\section{Collective excitations and Covariance}\label{App3}

To obtain physical insight, we consider the set of equations of motion \eqref{eq:LEoM} given in the main text. Their explicit solution is not possible analytically, but we can perform an eigenmode analysis. For this, 
we linearize the Eqs.\ \eqref{eq:LEoM} by considering small deviations
from the stationary state $(\a_0',\s_0,\S_0(\text{or }\zt_0))$ in the form of $\a'(t)=\a_0'+\delta\a'(t)$, $\s(t)=\s_0+\delta\s(t)$ and $\zt(t)=\zt_0+\delta\zt(t)$. The subsequent linearization  with respect to the deviations yields 
\begin{align}
  \delta\ddot\a' + 2\gam\delta\dot\a'+\omega_{\a'}^2
  \delta\a'&=c_1\delta\zt-c_2\delta\s\,,\nonumber\\
\delta\ddot\zt+\omega_\zt^2\delta\zt &= c_3 \delta\a'\,\label{eq:LEoM-st},\\
\delta\ddot\s + \omega_\s^2\delta\s&=-c_4 \delta\a'\nonumber\,,
\end{align}
with the frequencies $\omega_{\a'}^2=\Om^2+\gamma^2$, $\omega_\zt^2=4\oR{}V/(1-\S_0^2)$ and 
$\omega_\s^2=4\oR[3\oR/\s_0^4+Ve^{-\s_0^2}(1-2\s_0^2)/\sqrt{1-\S_0^2}+gN/\sqrt{2\pi}\s_0^3]$. 
Moreover, we have defined the coupling constants $c_1=2\sqrt{N}\la\Om{}\sqrt{1-\S_0^2}e^{-\s_0^2}$, 
$c_2=2\sqrt{N}\la\Om{}\s_0\S_0e^{-\s_0^2}$, $c_3=4\oR c_1 /\Om$, and $c_4=8\oR c_2 /\Om$.

In order to determine the collective excitation spectrum within a harmonic analysis, we define the rescaled quadrature operators $q_\mu = \epsilon_\mu \delta\mu$ and $p_\mu =\dot{q}_\mu/\omega_\mu$ with $\mu\in\{\a',\zt,\s\}$, and $\epsilon_{\a'}=1$, $\epsilon_{\zt}=\sqrt{c_1\omega_{\a'}/c_3\omega_\zt}$, $\epsilon_{\s}=\sqrt{c_2\omega_{\a'}/c_4\omega_{\s}}$.
Arranging the quadratures in a vector $\boldsymbol{x}=(q_{\a'},p_{\a'},q_{\zt},p_{\zt},q_\s,p_\s)^T$, the differential equation~\ref{eq:LEoM-st} can be written in the short form $\dot{\boldsymbol{x}}=\boldsymbol{M}\boldsymbol{x}$, with the matrix
\begin{equation}
\boldsymbol{M}=\begin{pmatrix}
0 & \omega_{\a'} & 0 & 0 & 0 & 0\\
-\omega_{\a'} & -2\gam & 2\lambda_{\a'\zt} & 0 & -2\lambda_{\a'\s} & 0\\
0 & 0 & 0 & \omega_\zt & 0 & 0\\
2\lambda_{\a'\zt} & 0 & -\omega_\zt & 0 & 0 & 0\\
0 & 0 & 0 & 0 & 0 & \omega_\s\\
-2\lambda_{\a'\s} & 0 & 0 & 0 & -\omega_\s & 0
\end{pmatrix}\,.
\end{equation}
Here, we have defined the effective coupling strengths $\la_{\a'\zt}=\sqrt{c_1 c_3/ 4\omega_{\a'} \omega_\zt}$ and $\la_{\a'\zt}=\sqrt{c_2 c_4/ 4\omega_{\a'} \omega_\zt}$.
Note that above the critical point $\la\geq\la_\cc$, both $c_1\sim\la^{-1}$ and $c_3\sim\la^{-1}$ are decreasing with $\lambda$ as $w_0\sim\lambda^{-2}$, from which follows that $\la_{\a'\zt}\rightarrow0$.
%
%
%
%

Below the critical point, the matrix~$\boldsymbol{M}$ becomes block-diagonal as $\la_{\a'\s}=0$.
In the limit of zero mechanical damping $\gamma \to 0$, this reduces to an analytically solvable eigenvalue problem of the matrix
\begin{equation}
\boldsymbol{M}'=\begin{pmatrix}
0 & \Om & 0 & 0\\
-\Om & 0 & 2\la_{\a'\zt} & 0\\
0 & 0 & 0 & \omega_\zt &\\
2\la_{\a'\zt} & 0 & -\omega_\zt & 0\\
\end{pmatrix}\,.
\end{equation}
$\boldsymbol{M}'$ has the four imaginary eigenvalues~$\nu_{i}$ given by the relation
$\nu^2 = - \left({\Om^2+\omega_\zt^2} \pm \sqrt{(\Om^2-\omega_\zt^2)^2 + 4 c_1 c_3}\right)/2$.
When $({\Om^2-\omega_\zt^2})^2\gg c_1 c_3$, which is well satisfied for the parameters chosen in this work, the eigenvalues can be estimated to
\begin{align}\begin{split}
\nu_{1,\pm} &= \pm i\Om\sqrt{1+(\la/\la_\cc)^2}\,,\\
\nu_{2,\pm} &= \pm i\omega_\zt\sqrt{1-(\la/\la_\cc)^2}\,,
\end{split}\end{align}
where $c_1 c_3/\Om^2\omega_\zt^2=(\la/\la_\cc)^2$ has been used.
The remaining two eigenvalues of the full matrix $\boldsymbol{M}$ are given by $\nu_{3,\pm}=\pm i \omega_\s$.

Moreover, for $\gamma=0$, the dynamics governed by $\dot{\boldsymbol{x}}=\boldsymbol{M}\boldsymbol{x}$ is generated by the Hamiltonian
\begin{align}
H &= \Om b_{\a'}^\dag b_{\a'}+\omega_\zt b^\dag_\zt b_\zt +\omega_\s b^\dag_\s b_\s \label{eq:ham2}\\
&+ \la_{\a'\zt}(b_{\a'}+ b_{\a'}^\dag)(b_{\zt}+ b_{\zt}^\dag)- \la_{\a'\s}(b_{\a'}+ b_{\a'}^\dag)(b_{\s}+ b_{\s}^\dag)\nonumber\,,
\end{align}
with bosonic algebra $[b_\mu,b^\dag_{\mu'}]=\delta_{\mu\mu'}$, where $b_\mu = (q_\mu + ip_\mu)/\sqrt{2}$.

The Hamiltonian \eqref{eq:ham2} is diagonalized by the Bogoliubov transformation
\begin{equation}
b_\mu = \sum_{k=1}^3 \Bigl[u^k_\mu d_k + \bigl(v^k_\mu\bigr)^* d_k^\dag\Bigr]\,,
\end{equation} 
with eigenenergies $\epsilon_k=\text{Im}(\nu_{k,+})$ and operators $d_k$, $d_k^\dag$ that satisfy bosonic commutation relations if we enforce
\begin{equation}
\sum_{\mu} \Bigl[u^k_\mu \bigl(u^k_\mu\bigr)^* - v^k_\mu \bigl(v^k_\mu\bigr)^*\Bigr] = 1\,.
\end{equation}
In order to estimate the entanglement between the condensate and membrane, we estimate the logarithmic negativity $E_\mathcal{N}$ from the ground state solution of \eqref{eq:ham2}.
The logarithmic negativity is related to the smallest symplectic eigenvalue of the reduced covariance matrix of the relevant modes. Therefore, we define the covariance matrix of the quadratures 
\begin{equation}
C_{kl}=\frac{1}{2}\langle x_kx_l + x_lx_k \rangle\,.
\end{equation}
To describe the entanglement between two different modes, the reduced covariance matrix $\boldsymbol{C'}$ is obtained by neglecting the columns and row of the irrelevant mode.
Then, the reduced covariance takes, in general, the form
\begin{equation}
\boldsymbol{C'} =  \begin{pmatrix}
\boldsymbol{U} & \boldsymbol{V}\\
\boldsymbol{V}^T & \boldsymbol{W}
\end{pmatrix}\,,
\end{equation}
and the logarithmic negativity can be expressed as~\cite{Adesso2004,Nagy2011,Arani2016}
\begin{equation}
E_\mathcal{N} = \text{max}\{0,-\text{log}(2\tilde{\nu}_-)\}\,,
\end{equation}
where
\begin{equation}
\tilde{\nu}_- = 2^{-1/2} \sqrt{\Sigma(\boldsymbol{C'})-\sqrt{\Sigma(\boldsymbol{C'})^2-4\text{det}\boldsymbol{C'}}}\,,
\end{equation}
and $\Sigma(\boldsymbol{C'})=\text{det}\boldsymbol{U}+\text{det}\boldsymbol{W}-2\text{det}\boldsymbol{V}$.
}
\section{Collective excitations in the Gross-Pitaevksii equation (GPE)}\label{App4}
\begin{figure}
\begin{center}
\includegraphics*[clip=true, trim=15 0 0 0, width=\columnwidth]{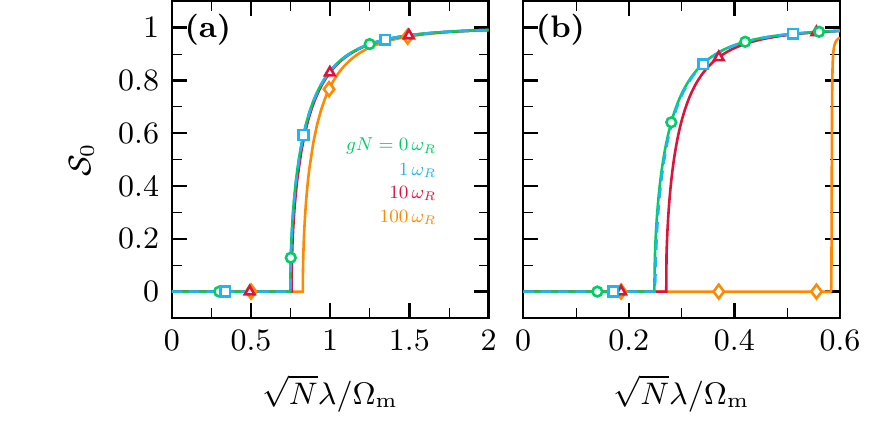} 
\caption{Order parameter $\S_0$ as a function of the coupling strength for (a) $V=200\,\oR$ and (b) $V=20\,\oR$. A different color corresponds to a different interaction strength $gN$, as indicated in (a). The other parameters are $\Om=100\,\oR$, $\gam=20\,\oR$. Here, we only show the results obtained by the GPE \maineqGPE \ of the main text.}\label{figS-int}
\end{center}
\end{figure}
A harmonic analysis can also be carried out within the GPE. The collective excitation spectrum follows 
by considering deviations from the stationary state $(\p_0,\a_0)$ in the form 
$\p(z,t)=e^{-i\mu{}t}[\p_0(z)+\delta\p(z,t)]$ and $\a(t)=\a_0+\delta\a(t)$. 
Without loss of generality, $\p_0(z)$ can be chosen to be real valued.
Then, linearizing the coupled equations~\maineqGPE~of the main text with respect to the deviations, 
the Bogoliubov-de Gennes equations 
\begin{align}
i\partial_t\delta\p(z) &= [h_0(z) + 2gN\p_0^2(z)]\delta\p(z) + gN\p_0^2(z)\delta\p^*(z)\nonumber\\
&-\la\sin(2z)\p_0(z)[\delta\a+\delta\a^*]\,,\label{eq:BdG}\\
i\partial_t \delta\a &= [\Om-i\gam]\delta\a - N\la \mathcal{Q}[\delta\p+\delta\p^*]\,,\nonumber
\end{align}
are obtained, with the linear operator $\mathcal{Q}[f]=\int dz\,\p_0(z)\sin(2z)f(z)$ and $h_0(z) = -\oR\partial_z^2 + V\sin^2(z)+Ng\p_0^2(z)-\la(\a_0+\a_0^*)\sin(2z)-\mu$.

\begin{figure}
\begin{center}
\includegraphics*[clip=true, trim=20 2 0 0, width=\columnwidth]{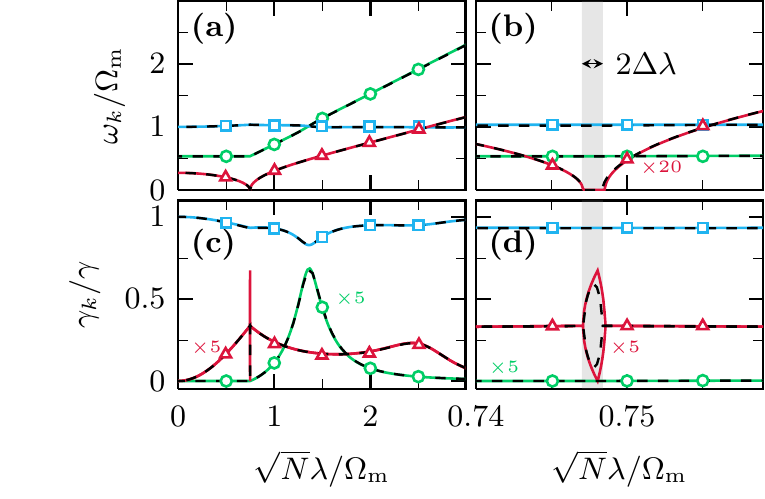}
\caption{(a), (b) Frequencies of collective excitations, and, (c), (d) decay rates of the collective eigenmodes (some curves are scaled by the factors indicated). Panels (b) and (d) show a zoom around $\la_\cc$
  with width $\Delta\la\simeq0.1\,\oR$.
  Different colors correspond to different eigenmodes. Curves in blue, green, and red correspond to the membrane mode, the condensate width mode, the atomic displacement excitation, correspondingly. The dashed curves show the GPE results. The parameters are $V=200\,\oR$, $\Om=100\,\oR$, $\gam=20\,\oR$ and $g=0$.
}\label{fig3old}
\end{center}
\end{figure}

The set of coupled differential equations~\eqref{eq:BdG} couples the deviations to their complex conjugates.
In that sense, the solutions are of the form $\delta\a(t)=\sum_k[e^{-i\nu_kt}\delta\a_{+,k}+e^{i\nu_kt}\delta\a_{-,k}^*]$ and $\delta\p(z,t)=\sum_k[e^{-i\nu_kt}u_{k}(z)+e^{i\nu_kt}v_k^*(z)]$ with complex frequencies $\nu_k$.
Within this ansatz, the frequencies are determined by solving the eigenvalue problem
\begin{equation}
\nu_k\boldsymbol{w}_k(z)=\boldsymbol{M}_{\rm GP}(z)\boldsymbol{w}_k(z)\,,
\end{equation}
with the vector $\boldsymbol{w}_k(z)=(\delta\a_{+,k}, u_k(z),\delta\a_{-,k},v_k(z))^T$ and the matrix
\begin{equation}
\boldsymbol{M}_{\rm GP}(z) = \begin{pmatrix} X & Y \\ -Y & -X
\end{pmatrix}
\end{equation}
where
\begin{align}\begin{split}
X=&\begin{pmatrix}
\Om-i\gam & - N\la\mathcal{Q}\\
-\la\sin(2z)\p_0 & h_0 +2gN\p_0^2
\end{pmatrix}\,,\\
Y=&\begin{pmatrix}
0 & - N\la\mathcal{Q}\\
-\la\sin(2z)\p_0 & gN\p_0^2
\end{pmatrix}\,.
\end{split}\end{align}
The eigenvalues of $\boldsymbol{M}_{\rm GP}(z)$ appear in pairs, such that if $\nu_k$ is an eigenvalue, the negative complex conjugate $-\nu_k^*$ is also an eigenvalue.
{The eigenfrequencies of the collective excitations without interatomic collisions are shown in 
Fig.\ \ref{fig3old}(a) and the corresponding decay rates in Fig.\ \ref{fig3old}(c) as a function of the atom-membrane coupling. A zoom around the critical value $\la_\cc$ is shown in Fig.\ \ref{fig3old}(b) and (d).}

{
\section{Experimental realization}\label{App5}
The predicted nonequilibrium quantum phase transition can be observed by available experimental set-ups \cite{Faber2017,Zhong2017}, as mentioned in the main text. For the benefit of the reader, we reproduce the relevant parameters of Ref.\ \cite{Zhong2017} here. The mass of the nanomembrane is $M=9.7 \times 10^{-11}$ kg, its eigenfrequency is $\Omega_m = 2 \pi \times 263.8$ kHz and the mechanical damping strength is found to be $\gamma = 2\pi \times 24.4 \times 10^{-3}$ Hz. A laser with $\omega_L = 2 \pi \times 384$ THz has been used and the optical cavity has a rather low finesse of $F=50$ to $120$. This justifies the adiabatic elimination of the light field. Then, a condensate of $^{87}$Rb atoms (atom mass $m = 86.9$ u $= 1.443\times 10^{-25}$ kg) has been used with $N=2 \times 10^6$ atoms. The recoil frequency then amounts to $\omega_R=2 \pi \times  3.8$ kHz. Regarding the possible values of the s-wave interaction strength, $\omega_\perp = 2 \pi \times 62 $ to $2 \pi \times 85$ Hz can be realized. 
Likewise, the scattering length is  $a_s = 95$ to $98$ $a_{\rm Bohr}$. This amounts to a one-dimensional effective interaction strength of $g_{1D}/\hbar = 2\omega_\perp a_s = 2 \pi \times (11.4 \dots 16.6)$ kHz $a_{\rm Bohr}$.
}
\begin{figure}[b]
\begin{center}
\includegraphics*[clip=true, trim=20 2 0 0, width=\columnwidth]{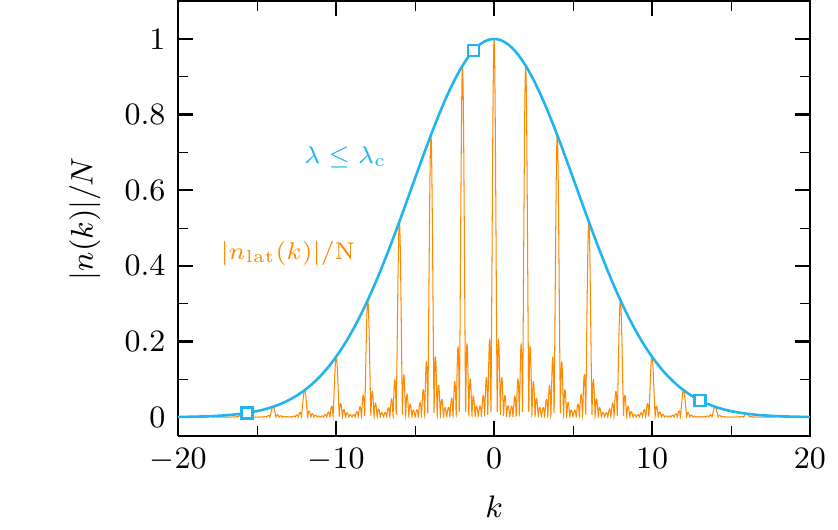}
\caption{Momentum distributions of atoms in a single lattice site $\abs{n(k)}$, and of the full lattice $n_\text{lat}(k)$ for $M=10$ sites below the critical coupling $\la_\cc$. The parameters are $V=200\,\oR$, $\Om=100\,\oR$, $\gam=20\,\oR$ and $g=0$.}\label{figS1}
\end{center}
\end{figure}
{
\section{Role of atom-atom interaction}\label{App6}
In the main text, we have shown the results for the case $gN=0$ of non-interacting atoms.
In fact, current experimental set-ups achieve an effective one-dimensional interaction strength in the region $gN\sim1-10\,\oR$ by confining the atomic motion within a harmonic potential with a horizontal trapping frequency of about $100$ Hz and assuming  $\sim 10^4$ atoms in a single potential well.

In Fig.~\ref{figS-int}, we show the order parameter for different values of the atom-atom interaction strength and two values of the potential depth (a) $V=200\,\oR$ and (b) $V=20\,\oR$.
In both cases, we observe no qualitative difference in the order parameter between the non-interacting case $g=0$ and the interacting case (in the mean-field sense) with $gN\ll V$. Finally, we note that the conclusion may be different in the regime of strong interaction where Mott physics in the lattice occurs in addition. 
}
\section{Momentum Distribution of the atoms}\label{App7}
The momentum distribution $n(k)$ of the atoms in a single lattice site is shown in Fig.\ 4(b) of the main text. 
From the single site Gaussian ansatz, it follows that
\begin{equation}
n(k)=N\int dz e^{ik(z-\zt_0)} \abs{\psi(z)}^2=Ne^{-(k\s_0/2)^2}\,.
\end{equation}
If we also take into account that we have, in fact, a periodic potential with $M$ lattice sites, the momentum distribution $n_\text{lat}(k)$ of the atoms can be calculated as follows. We start from the atomic density in the lattice
\begin{equation}
 n_\text{lat}(z) \simeq \frac{N}{M}\sum_{l=0}^{M-1}\frac{1}{\sqrt{\pi}\s_0}e^{-{(z-\zt_{0}-\pi l)^2}/{\s_0^2}} \, ,
\end{equation}
where $M$ denotes the number of lattice sites (potential wells). The corresponding momentum distribution of the atoms in the lattice is then
\begin{equation}
n_\text{lat}(k) = \int dz e^{ikz} n_\text{lat}(z) \equiv f(k) n(k)e^{ik\zt_0}
\end{equation}
with the form factor
\begin{equation}
\abs{f(k)} = \frac{\sin(\pi Mk/2)}{M\sin(\pi k/2)}\, .
\end{equation}
The atomic momentum distribution $n_\text{lat}(k)$ of the whole lattice is shown in Fig.\ \ref{figS1} together with the single site momentum distribution $n(k)$.

\end{document}